%% file: main.tex
\documentclass[11pt]{article}
\usepackage{amsmath}

\newcommand{\gi}[2]{\gamma^{i_{#1}\cdots i_{#2}}}
\newcommand{\gj}[2]{\gamma_{j_{#1}\cdots j_{#2}}}
\newcommand{\comm}[3]{\left[ #1, #2 \right]_{#3}}

\begin{document}
\begin{flushright}
NA-DSF-38/2007
\end{flushright}
\vspace{1ex}

\begin{center}
\textbf{\Large General (anti-)commutators of gamma matrices}\\[4ex]%
Wolfgang M\"uck\\[3ex]%
\textit{Dipartimento di Scienze Fisiche, Universit\`a degli Studi di
  Napoli ``Federico II''\\ and INFN, Sezione di Napoli --- via Cintia, 80126 Napoli, Italy}\\%
E-mail: \texttt{mueck@na.infn.it}
\end{center}
\vspace{1ex}
\begin{abstract}
Commutators and anticommutators of gamma matrices with arbitrary numbers of (antisymmetrized) indices are derived.
\end{abstract}

\vfill

\input{gamma}

This research was supported by the European Commission, project MRTN-CT-2004-005104, and by the MiUR-COFIN project 2005-023102.

\input{main.bbl}

\end{document}

%% file: gamma.tex
Gamma matrix algebra is ubiquitous in many calculations in high energy physics. Whereas it is a fairly simple business in four dimensions, in higher-dimensional applications such as supergravity or M-theory, it becomes quite involved, because the number of independent matrices grows quickly with the number of space-time dimensions. To ease such calculations, on the one hand, one can resort to the help of computer algebra packages \cite[and references therein]{Gran:2001yh}. On the other hand, one can look in the literature for reference tables, such as the appendix of \cite{Candelas:1984yd}, where the commutators and anti-commutators of gamma matrices with up to four indices are listed. The table in \cite{Candelas:1984yd} is, to my knowledge, the most complete such list, but it contains typographical errors, as has been noticed in \cite{Frey:2004rn}. The purpose of this short note is to derive general formulae for the commutators and anti-commutators of gamma matrices with any number of indices. A general treatment is possible, because the (anti-)commutators do not depend on the space-time dimension, $d$, except for the fact that the number of indices any gamma matrix can carry is limited by $d$. The final formulae take the form of explicit sums and do not involve recursion relations.

Consider the $d$-dimensional Clifford algebra generated by the matrices $\gamma^i$ $(i=1,\ldots,d)$, which satisfy
\begin{equation}
\label{Clifford}
  \gamma^i \gamma^j +\gamma^j \gamma^i = 2g^{ij}~,
\end{equation}
where $g^{ij}$ is the inverse metric tensor. Throughout this paper, indices shall be raised and lowered with $g^{ij}$ and $g_{ij}$, respectively. Note that the metric $g_{ij}$ can be curved or flat, and also its signature will be irrelevant for what follows. A useful basis of the Clifford algebra of matrices is given by the antisymmetrized products of the $\gamma^i$,\footnote{The antisymmetrization includes a factor $1/k!$ for normalization. For odd $d$, the basis such defined is overcomplete, but this will not influence our analysis.}
\begin{equation}
\label{basis}
  \gi{1}{k} = \gamma^{[i_1} \gamma^{i_2} \cdots \gamma^{i_k]} \qquad (1\leq k \leq d)~,
\end{equation}
and by the identity matrix, which we may include in the notation \eqref{basis} by allowing also for $k=0$.  We will formally allow also for $k>d$ implying that the corresponding matrix vanishes due to the antisymmetrization of the indices. 

We shall obtain general formulae for all commutators and anti-commutators of the $\gamma$-matrices \eqref{basis}. A useful notation we will employ is the generalized commutator bracket
\begin{equation}
\label{comm.def}
  \comm{a}{b}{x} = ab + xba~,
\end{equation}
where $x=\pm 1$. Our convention for the generalized Kronecker delta symbol is
\begin{equation}
 \label{Kronecker}
  \delta^{i_1\cdots i_k}_{j_1\cdots j_k} = \delta^{[i_1}_{j_1} \cdots \delta^{i_k]}_{j_k}~.
\end{equation}

Let us start with the easiest piece and write
\begin{align}
\notag
  \gamma_j \gi{1}{k} &= 
  \gamma_j \gamma^{[i_1} \cdots \gamma^{i_k]} \\
\notag 
  &= - \gamma^{[i_1} \gamma_j \gamma^{i_2} \cdots \gamma^{i_k]} 
  + 2 \delta_j^{[i_1} \gamma^{i_2} \cdots \gamma^{i_k]}~.
\end{align}
After pulling $\gamma_j$ through the other matrices, we end up
with
\begin{equation}
\label{com.1k-}
  \comm{\gamma_j}{\gi{1}{k}}{(-1)^{k+1}}
  = 2 k \delta_j^{[i_1} \gamma^{i_2\cdots i_k]}~.
\end{equation}
This is a commutator for even $k$ and an anti-commutator for
odd $k$. Finding the other bracket (anti-commutator for even $k$,
commutator for odd $k$) is best done using induction.
Let us assume that, for some $k$, the following relation holds:
\begin{equation}
\label{com.1k+}
  \comm{\gamma_j}{\gi{1}{k}}{(-1)^k}
  = 2 \gamma_j{}^{i_1 \cdots i_k}~.
\end{equation}
Consider $\gamma_j{}^{i_1\cdots i_{k+1}}$ and rewrite it as 
\begin{equation}
\label{ind.k1}
  2 \gamma_j{}^{i_1\cdots i_{k+1}} = \frac2{k+2} \left( \gamma_j
  \gi{1}{k+1}
  -(k+1) \gamma^{[i_1} \gamma_j{}^{i_2\cdots i_{k+1}]} \right)~.
\end{equation}
Applying the hypothesis \eqref{com.1k+} on the second term in the parentheses and using then \eqref{Clifford} and \eqref{com.1k-}, one obtains after a bit of algebra
\begin{equation}
\label{ind.k2}
  2 \gamma_j{}^{i_1\cdots i_{k+1}} = \gamma_j \gi{1}{k+1} +
  (-1)^{k+1} \gi{1}{k+1} \gamma_j
  = \comm{\gamma_j}{\gi{1}{k+1}}{(-1)^{k+1}}~.
\end{equation}
Thus, if the hypothesis \eqref{com.1k+} is valid for some $k$, then it will also hold for $k+1$. Therefore, as \eqref{com.1k+} holds for $k=1$ by the definition of $\gamma^{ij}$, we have shown that it holds for any $k$.

After this little exercise, we are ready to face the general cases $\comm{\gj{1}{l}}{\gi{1}{k}}{\pm}$. Our hypotheses, which we shall prove again by induction, are the following:
\begin{align}
\label{com.kl+}
  \comm{\gj{1}{l}}{\gi{1}{k}}{(-1)^{kl}}
  &= 2 \sum\limits_{m=0}^{\infty} (-1)^m
       (2m)! \binom{k}{2m} \binom{l}{2m}
       \delta^{[i_1\cdots i_{2m}}_{[j_1 \cdots j_{2m}} 
       \gamma_{j_{2m+1} \cdots j_l]}^{}{}^{i_{2m+1} \cdots i_k]}_{}~,\\
\notag 
  \comm{\gj{1}{l}}{\gi{1}{k}}{(-1)^{kl+1}}
  &= 2 \sum\limits_{m=0}^{\infty} (-1)^{m+l+1}
       (2m+1)! \binom{k}{2m+1} \binom{l}{2m+1} \\
\label{com.kl-}
  &\quad \times
       \delta^{[i_1\cdots i_{2m+1}}_{[j_1 \cdots j_{2m+1}} 
       \gamma_{j_{2m+2} \cdots j_l]}^{}{}^{i_{2m+2} \cdots i_k]}_{}~.
\end{align}
Notice that the sums in these formulae are actually not infinite, but they terminate, because the binomial coefficients vanish for large enough $m$. Similarly, we could have formally extended the sums to $-\infty$. It is straightforward to verify that \eqref{com.kl+} and \eqref{com.kl-} reduce to \eqref{com.1k+} and \eqref{com.1k-}, respectively, if $l=1$.

The proof by induction can be based on the identity
\begin{equation}
 \label{comm.gen}
\begin{split}
   \comm{\gj{1}{l+1}}{\gi{1}{k}}{x} &=
  \frac12 \comm{\gamma_{j_{[1}}}{\comm{\gj{2}{l+1]}}{\gi{1}{k}}{x(-1)^k}}{(-1)^{k+l}}
\\ & \quad 
 +\frac12 \comm{\gj{[1}{l}}{\comm{\gamma_{j_{l+1]}}}{\gi{1}{k}}{(-1)^{k+1}}}{x(-1)^{k+l+1}}~,
\end{split}
\end{equation}
which holds for $x^2=1$. Let us assume that \eqref{com.kl+} and \eqref{com.kl-} hold for some $l$ and any $k$. Choosing $x=(-1)^{k(l+1)}$, we use \eqref{com.kl+} and \eqref{com.1k+} in the first term on the right hand side of \eqref{comm.gen}, and \eqref{com.1k-} and \eqref{com.kl-} in the second term. Combining both terms (after shifting the summation index $m$ by one in the second term), we end up with \eqref{com.kl+} with $(l+1)$ in place of $l$. Similarly, choosing $x=(-1)^{k(l+1)+1}$, we use \eqref{com.kl-} and \eqref{com.1k+} in the first term, \eqref{com.1k-} and \eqref{com.kl+} in the second term and obtain in the end \eqref{com.kl-} with $l+1$ in place of $l$. Therefore, as \eqref{com.kl+} and \eqref{com.kl-} hold for $l=1$ and any $k$, we have shown that they hold for any $k$ and $l$.

Equations \eqref{com.kl+} and \eqref{com.kl-} are the results of this paper. We leave it as an exercise for the reader to find three typographical errors in the list of (anti-)commutators given in the appendix of \cite{Candelas:1984yd}. (In addition, a term containing a gamma matrix with eight indices has been omitted in the penultimate formula of this list.)

%% file: main.bbl
\providecommand{\href}[2]{#2}\begingroup\raggedright\endgroup